\documentclass[12pt]{article}
\usepackage{amsfonts}
\usepackage{amsmath}
\usepackage{amssymb}
\usepackage{mathptmx}
\usepackage{geometry}
\usepackage{graphicx}
\usepackage{hyperref}
\usepackage{cancel}
\usepackage{textcomp}
\usepackage{tikz}
\usepackage{bm}
  \usepackage{mathrsfs}
  \usepackage{filecontents}
\usepackage{times}
\usepackage{epsfig}
\usepackage{xcolor}
\usepackage{slashed}
\usepackage{booktabs}
\usepackage{latexsym}
\usepackage{verbatim}
\usepackage{extarrows}
\usepackage{multirow}
\usepackage{rotating}
\usepackage{colortbl}
\usepackage{indentfirst}
\usepackage{float}
\definecolor{mygray}{gray}{.9}
\definecolor{intnull}{RGB}{213,229,255}
\usepackage{arydshln}
\usepackage{diagbox}
\usepackage{makecell}
\usepackage{array}
\usepackage{cite}

\usepackage{caption}
\usepackage{tikz}
\usetikzlibrary{arrows,shapes,chains}
\usepackage{graphicx, subfig}
\begin{document}
\renewcommand{\thefootnote}{\fnsymbol{footnote}}
\baselineskip=16pt
\pagenumbering{arabic}
\vspace{1.0cm}
\begin{center}
{\Large\sf High-dimensional Schwarzschild black holes in scalar-tensor-vector gravity theory}
\\[10pt]
\vspace{.5 cm}

{Xin-Chang Cai\footnote{E-mail address: caixc@mail.nankai.edu.cn} and Yan-Gang Miao\footnote{Corresponding author. E-mail address: miaoyg@nankai.edu.cn}}
\vspace{3mm}

{School of Physics, Nankai University, Tianjin 300071, China}

\vspace{4.0ex}

{\bf Abstract}
\end{center}

We obtain a high-dimensional Schwarzschild black hole solution in the scalar-tensor-vector gravity (STVG), and then analyze the influence of parameter $\alpha$ associated with a deviation of the STVG theory from General Relativity on event horizons and Hawking temperature. We calculate the quasinormal mode frequencies of massless scalar field perturbations for the high-dimensional Schwarzschild STVG black hole by using the sixth-order WKB approximation method and the unstable null geodesic method in the eikonal limit. The results show that the increase of parameter $\alpha$ makes the scalar waves decay slowly, while the increase of the spacetime dimension makes the scalar waves decay fast. In addition, we study the influence of parameter $\alpha$ on the shadow radius of this high-dimensional Schwarzschild STVG black hole and find that the increase of parameter $\alpha$ makes the black hole shadow radius increase, but the increase of the spacetime dimension makes the black hole shadow radius decrease.  Finally, we investigate the energy emission rate of the high-dimensional Schwarzschild STVG black hole, and find that the increase of parameter $\alpha$ makes the evaporation process slow, while the increase of the spacetime dimension makes the process fast.

\newpage

\section{Introduction}

The gravitational waves from a binary black hole merger detected by LIGO and Virgo collaborations~\cite{P3,P4} and the first black hole image from the supermassive black hole in the center of galaxy M87 detected by the Event Horizon Telescope (EHT) collaboration~\cite{PP1,PP2} have greatly stimulated our enthusiasm in black hole physics. The former provides a new window for the study of black hole perturbations, while the latter promotes the research of black hole shadows. Up to now, the perturbations and shadows of various black hole models have been analyzed extensively and deeply in Einstein's gravity and other modified gravity theories, see, for example, some references\cite{PC1,PC2,PC3,PC4,PGA2,PGA3,PC5,PC6,PC7,PC8}.

As is known,  Einstein gravitational theory  has achieved great success in the past century and has been tested by numerous experiments, but it cannot explain~\cite{P19,P20,P21,P22} the difference between the dynamics of galaxies and the amount of luminous matter contained in the galaxies. One possible solution is to assume the existence of   exotic dark matter, which, unfortunately, has not yet been detected directly. The other solution is to modify~\cite{P23} the law of gravity on a scale that has not been widely tested by Newtonian gravity or general relativity. Among various modified gravitational theories, a good candidate is the scalar-tensor-vector gravity (STVG) theory~\cite{P1}. In addition to the metric tensor field, a scalar field and a massive vector field are also introduced in this theory, where the former can strengthen the gravitational attraction  and the latter can generate an effective repulsive gravitational force. The STVG theory not only describes~\cite{P24,P25,P26,P27} the dynamics of galaxies without the assumption of existence of dark matter in the universe, but also explains~\cite{P28,P29} the solar system, the growth of structure, the cosmic microwave background (CMB) acoustical power spectrum data, and so on. Recently, the quasinormal modes of electromagnetic and gravitational perturbations and black hole shadows in the STVG theory have been made progress~\cite{P30,P32,P31,PC12,PP3,P36,P33}.

In Einstein's gravity, besides the Schwarzschild black hole in the four-dimensional spacetime, there exists its  high-dimensional extension --- the Schwarzschild-Tangherlini black hole~\cite{P18}. Thus, we construct the high-dimensional Schwarzschild STVG black hole and study its quasinormal modes and shadows. Our main motivations for this high-dimensional extension are as follows.

\begin{itemize}

\item The string theory predicts~\cite{P34} the existence of extra dimensions, and the scenarios involved in large extra dimensions and TeV-scale gravity make the production of high-dimensional black holes in future colliders become a conceivable possibility~\cite{PVT2,PVT3}. As a result, high-dimensional black holes have been paid a lot of attentions.  Therefore, it is necessary to know whether a high-dimensional STVG black hole has some characteristics different from those the four-dimensional one has.

\item The AdS/CFT correspondence has been widely used~\cite{P35} in various fields of physics in recent years, which brought about interest in high-dimensional black holes. Especially, a weakly coupled gravitational theory in the five-dimensional AdS spacetime corresponds to a strongly-coupled four-dimensional gauge field theory. The STVG theory is regarded as a promising alternative to general relativity in four dimensions and has good achievements in the interpretation of the solar system observations, the rotation curves of galaxies, and the dynamics of galactic clusters, etc. If this theory could be combined with the AdS/CFT correspondence and then applied to particle physics, e.g., in the strongly coupled hot quark-gluon plasma~\cite{PVT1}, its importance would be greatly enhanced. Such a combination will be an extension of fields of physics for this modified gravity theory. Therefore, it is necessary to construct a high-dimensional STVG black hole in advance to pave the way for subsequent researches.

\item Based on the experimental background that both gravitational waves and black hole shadows have been detected, it is of great significance to study the quasinormal modes and black hole shadows of the high-dimensional Schwarzschild STVG black hole and compare them with those of the Schwarzschild-Tangherlini black hole~\cite{P13,PP4} of Einstein's gravity.

\item The high-dimensional Schwarzschild STVG black hole we constructed is not only an attempt to extend the STVG theory to a high-dimensional spacetime, but also fills the gap between the counterpart of this theory and the Schwarzschild-Tangherlini black hole under Einstein's theory.

\end{itemize}

The paper is organized as follows. In Sec. 2, we briefly introduce the field equations of the STVG theory and give the high-dimensional Schwarzschild STVG black hole solution. We analyze the event horizons and Hawking temperature of this black hole model in Sec. 3. In Sec. 4, we calculate the quasinormal mode frequencies of massless scalar field perturbations for the high-dimensional Schwarzschild STVG black hole by using the sixth-order WKB approximation method and the unstable null geodesic method in the eikonal limit as well. In Sec. 5,  we investigate the  black hole  shadow  radius  and the energy emission rate of the  high-dimensional  Schwarzschild  STVG black hole. Finally, we make a simple summary in Sec. 6. We use the units $c=G_{\rm N}=k_{\rm B}=\hbar=1$ and the sign convention $(-,+,+,+)$ throughout the paper.

\section{Scalar-tensor-vector gravity and its high-dimensional Schwarz-schild black hole}

The action of the STVG  theory  can be written as~\cite{P1},
\begin{equation}
\label{1}
S=S_{\rm GR}+S_{\phi}+S_{\rm S}+S_{\rm M},
\end{equation}
where $S_{\rm GR}$, $S_{\phi}$, and $S_{\rm S}$ stand for the Einstein-Hilbert action, the action of a massive vector field, and the action of a scalar field, respectively,
\begin{equation}
\label{2}
S_{\rm GR}=\frac{1}{16\pi }\int d^{D}x\sqrt{-g}\frac{1}{G}R,
\end{equation}
\begin{equation}
\label{3}
S_{\phi}=-\frac{1}{4\pi }\int d^{D}x\sqrt{-g}\left(K-\frac{1}{2}\tilde{\mu}^{2}\phi ^{\mu }\phi _{\mu}\right),
\end{equation}
\begin{equation}
\label{4}
S_{\rm S}=\int d^{D}x\sqrt{-g}\left[\frac{1}{G^{3}}\left(\frac{1}{2}g^{\mu \nu }\nabla _{\mu }G \nabla_{\nu  }G-V_G(G)\right)+\frac{1}{\tilde{\mu}^{2}G}\left(\frac{1}{2}g^{\mu \nu }\nabla _{\mu }\tilde{\mu} \nabla_{\nu  }\tilde{\mu}-V_{\tilde{\mu}}(\tilde{\mu})\right)\right],
\end{equation}
and $S_{\rm M}$ denotes the action of possible matter sources. This action is the direct extension of the four-dimensional one to a higher-dimensional spacetime. In Eqs.~(\ref{2})-(\ref{4}), $D$ represents the dimension of spacetimes,  $\phi^{\mu}$ a Proca-type massive vector field with mass $\tilde{\mu}$, $K$ the kinetic term of the vector field $\phi^{\mu}$, $G(x)$ and $\tilde{\mu}(x)$ two scalar fields that vary with respect to time and space, and $V_G(G)$ and $V_{\tilde{\mu}}(\tilde{\mu})$ their corresponding self-interaction potentials which can be set to be zero for simplicity in getting the black hole solution~\cite{P24}.  Moreover, $K$ is usually chosen as $K=\frac{1}{4}B^{\mu \nu}B_{\mu \nu}$ with the linear tensor field $B_{\mu\nu}=\partial_{\mu}\phi_{\nu}-\partial_{\nu}\phi_{\mu}$.

When solving field equations for a black hole solution, we can ignore the effect of the mass $\tilde{\mu}$ of the vector field $\phi^{\mu}$ according to Ref.~\cite{P2}. In addition, we treat $G$ as a constant dependent on $\alpha$, i.e.,
\begin{equation}
G=G_{\rm N}(1+\alpha),
\end{equation}
where $G_{\rm N}$ denotes Newton's gravitational constant in the $D$-dimensional black hole spacetime  and $\alpha$ is a dimensionless parameter.  Since the STVG theory returns to Einstein's general relativity (GR) for the case of $\alpha =0$, we can regard $\alpha $ as a deviation parameter of the STVG theory from GR. Therefore, we can simplify Eq.~(\ref{1}) to the following form for the vacuum solution,
\begin{equation}
\label{5}
S=\frac{1}{16\pi }\int d^{D}x\sqrt{-g}\left(\frac{R}{ G }-B^{\mu \nu}B_{\mu \nu}\right).
\end{equation}

By varying  the action Eq.~(\ref{5}) with respect to $g_{\mu \nu}$, one has the field equations,
\begin{equation}
\label{6}
G_{\mu \nu }=-8\pi GT_{\mu \nu }^{\phi },
\end{equation}
where $G_{\mu \nu}$ is the Einstein tensor, and the energy-momentum tensor $T_{\mu \nu}^{\phi}$ for the vector field $\phi ^{\mu}$ takes the form,
\begin{equation}
\label{7}
T_{\mu \nu}^{\phi}=-\frac{1}{4\pi}\left({B_{\mu}}^{\sigma}B_{\nu \sigma}-\frac{1}{4}g_{\mu \nu}B^{\rho \sigma}B_{\rho \sigma}\right).
\end{equation}
By varying the action Eq.~(\ref{5}) with respect to the vector field $\phi^{\mu}$, one obtains the field equations,
\begin{equation}
\label{8}
\nabla _{\nu  }B^{\mu \nu }=0,
\end{equation}
\begin{equation}
\label{9}
\nabla _{\sigma }B_{\mu \nu }+\nabla _{\mu  }B_{ \nu\sigma  }+\nabla _{\nu }B_{\sigma \mu  }=0.
\end{equation}

In order to gain the solution of a high-dimensional  static spherically symmetric black hole in the STVG, one usually assumes the following form for the line element,
\begin{equation}
\label{10}
ds^{2}=-f(r)dt^{2}+\frac{dr^{2}}{f(r)}+r^{2}d\Omega ^{2}_{D-2},
\end{equation}
where $d\Omega^{2}_{D-2}=d\chi^{2}_{2}+\prod_{i=2}^{D-2}\mathrm{sin}^{2}\chi_{i}d\chi_{i+1}^{2}$ represents the line element on the $(D-2)$-dimensional unit sphere.
Comparing Eqs.~(\ref{6}), (\ref{8}), and (\ref{9}) with those of the high-dimensional Reissner-Nordstr$\mathrm{\ddot{o}}$m black holes in Einstein gravity~\cite{P9,P10}, we find the similarity between them. As a result, we deduce the metric function $f(r)$ of the high-dimensional Schwarzschild STVG black hole solution and the vector field, respectively, by considering the correspondence~\cite{P2,P33} between the electric charge in the high-dimensional Reissner-Nordstr$\mathrm{\ddot{o}}$m black hole and the gravitational source charge of the vector field $\phi^{\mu}$,
\begin{equation}
\label{13}
f(r)=1-\frac{m}{r^{D-3}}+\frac{Gq^{2}}{r^{2(D-3)}},
\end{equation}
\begin{equation}
\label{16}
\phi _{\mu }=\left(\frac{4\pi \sqrt{\alpha G_{N}}M}{(D-3)\Omega _{D-2}r^{D-3}},0,0,0\right),
\end{equation}
where $m$ and $q$ are defined by
\begin{equation}
\label{14}
m\equiv \frac{16\pi GM}{(D-2)\Omega _{D-2}},\qquad q\equiv \frac{8\pi \sqrt{\alpha G_{N}}M}{\sqrt{2(D-2)(D-3)}\Omega _{D-2}},
\end{equation}
and $\Omega_{D-2}=\frac{2\pi^{\frac{D-1}{2}}}{\Gamma (\frac{D-1}{2})}$ means the area of a $(D-2)$-dimensional unit sphere.
It is obvious that the metric function Eq.~(\ref{13}) turns back to the Schwarzschild STVG black hole solution given by Moffat~\cite{P2} for the case of $D=4$.

Similar to the high-dimensional Reissner-Nordstr\"om  black hole of Einstein's GR, the high-dimensional Schwarzschild STVG black hole has two horizons,
\begin{equation}
\label{18}
r_{\pm }=\left(\frac{m}{2}\pm\frac{\sqrt{m^{2}-4Gq^{2}}}{2}\right)^{\frac{1}{D-3}},
\end{equation}
where $r_{-}$ is the inner horizon called the Cauchy horizon and $r_{+}$ is the outer horizon called the event horizon. In addition, when $\alpha =0$, the two horizons are merged into the event horizon of the Schwarzschild-Tangherlini black hole~\cite{P18,P37}. We note that the four-dimensional charged Reissner-Nordstr\"om STVG black hole solution~\cite{PGA1} is different from the four-dimensional Schwarzschild STVG black hole solution given by Moffat~\cite{P2}. So, our solution, see Eqs.~(\ref{10}), (\ref{13}), (\ref{16}), and (\ref{14}), should be regarded as the high-dimensional Schwarzschild black hole solution in the STVG theory. Next, we shall only compare it with the Schwarzschild-Tangherlini black hole in Einstein gravity.

\section{Horizon and Hawking temperature of high-dimensional \\Schwarzschild STVG black holes}

By solving $f(r_{\rm H})=0$ in Eq.~(\ref{13}), we can get the relationship between the black hole mass $M$ and the event horizon radius $r_{\rm H}$ as follows,
\begin{equation}
\label{19}
M=\frac{r_{\rm H}^{D-3} \left(A-\sqrt{A^2-4 B^2 G}\right)}{2 B^2 G},
\end{equation}
where the coefficients $A$ and $B$ are defined by
\begin{equation}
\label{20}
A\equiv \frac{16 \pi G}{(D-2) \Omega_{D-2}},  \qquad    B\equiv \frac{8 \pi  \sqrt{\alpha G_{\rm N}}}{\sqrt{2(D-2) (D-3)} \Omega_{D-2}}.
\end{equation}
Then, we obtain the Hawking temperature $T_{\rm H}$ on the event horizon,
\begin{equation}
\label{21}
T_{\rm H}=\frac{f'(r_{\rm H})}{4\pi}=\frac{(D-3)\left(A \sqrt{A^2-4 B^2 G}-A^2+4 B^2 G\right)}{8 \pi  B^2 G r_{\rm H}}.
\end{equation}

We draw the graph of the metric function $f(r)$ with respect to $r$ for different values of $\alpha$ in spacetime dimension $D=4, 5, 6, 7$ in Fig. 1, where the case of $D=4$~\cite{P2} is attached for comparison. It should be noted that the parameter $\alpha=0$ represents the case of the Schwarzschild-Tangherlini black hole of Einstein's gravity. From this figure, we can see that the event horizon radius $r_{\rm H}$ increases with the increase of the parameter $\alpha$ when $D=5, 6, 7$, which is similar to the situation of the four-dimensional Schwarzschild STVG black hole. In addition, we can also see that the high-dimensional Schwarzschild STVG black hole always has one more horizon than the Schwarzschild-Tangherlini black hole before reaching the extreme case.

We plot the graph of the Hawking temperature $T_{\rm H}$ with respect to $r_{\rm H}$ for different values of $\alpha$ in spacetime dimension $D=4, 5, 6, 7$ in Fig. 2, where the case of $D=4$~\cite{P2} is attached for comparison. It should be noted that the parameter $\alpha=0$ represents the case of the Schwarzschild-Tangherlini black hole of Einstein's gravity.  From this figure, we can find that the relationship between the Hawking temperature of the high-dimensional Schwarzschild STVG black hole and the event horizon radius is similar to that of the Schwarzschild-Tangherlini black hole, that is, the Hawking temperature $T_{\rm H}$ decreases monotonically with the increase of the event horizon radius $r_{\rm H}$ when $D=5, 6, 7$. In addition, we can also see that the Hawking temperature $T_{\rm H}$ decreases with the increase of parameter $\alpha$ in the high-dimensional Schwarzschild STVG black hole, which is similar to the situation of the four-dimensional Schwarzschild STVG black hole.

\begin{figure}[h]
		\centering
		\begin{minipage}{.5\textwidth}
			\centering
			\includegraphics[width=80mm]{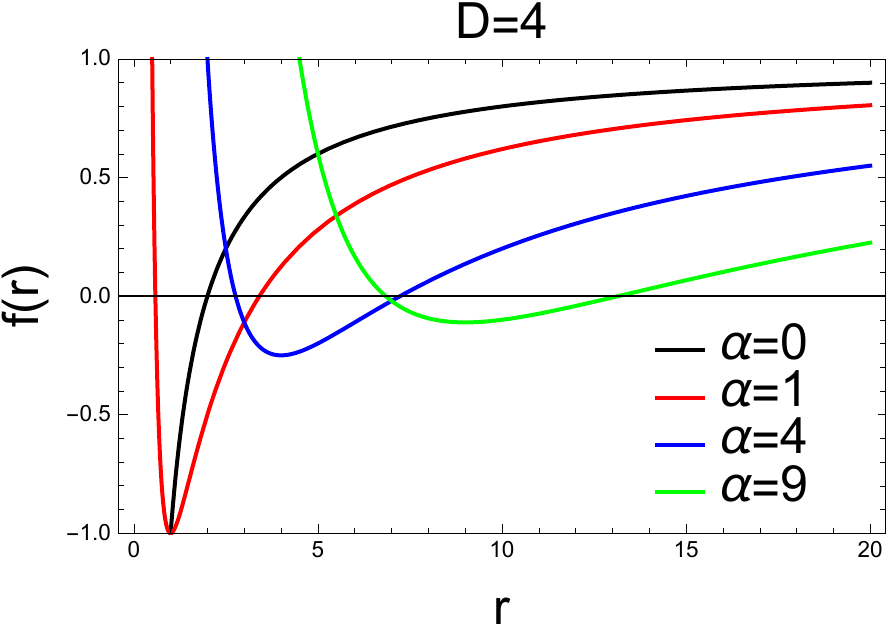}
		\end{minipage}%
		\begin{minipage}{.5\textwidth}
			\centering
			\includegraphics[width=80mm]{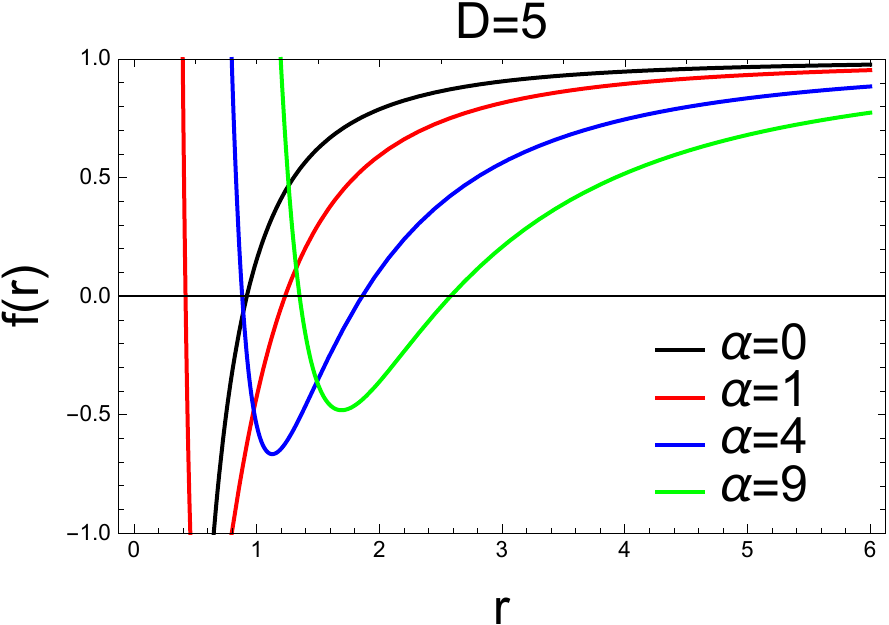}
		\end{minipage}
		\centering
		\begin{minipage}{.5\textwidth}
			\centering
			\includegraphics[width=80mm]{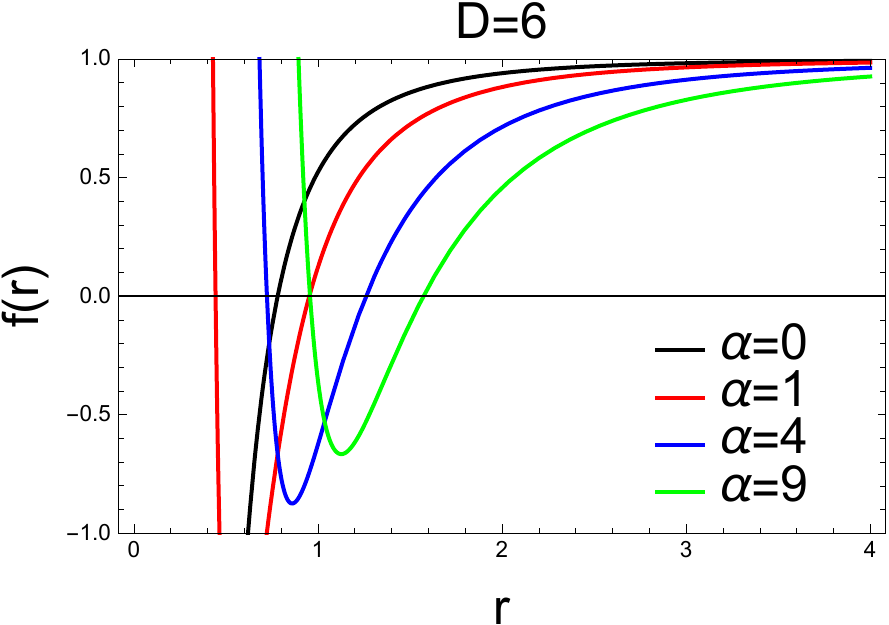}
		\end{minipage}%
		\begin{minipage}{.5\textwidth}
			\centering
			\includegraphics[width=80mm]{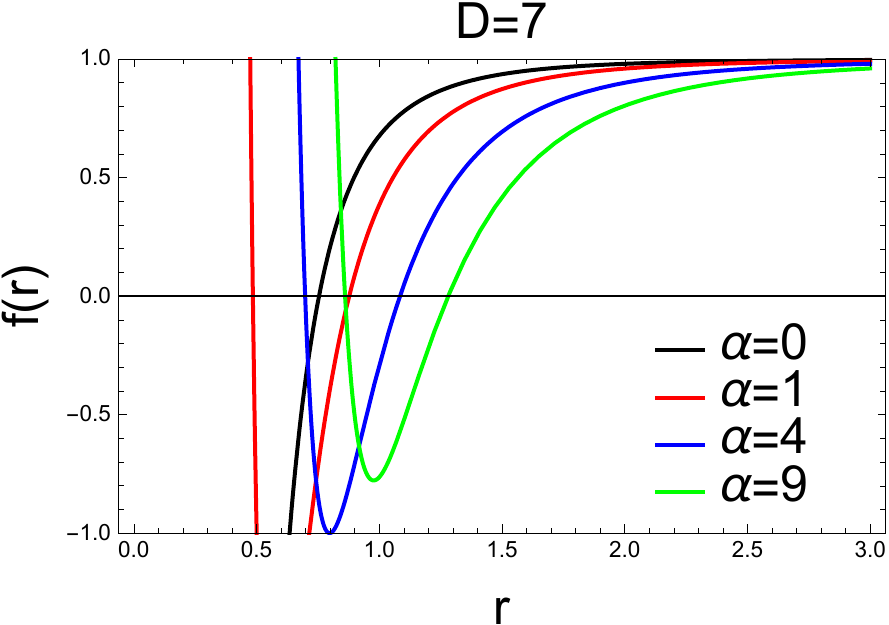}
		\end{minipage}
		\caption*{Fig. 1. Function $f(r)$ with respect to $r$ for different values of $\alpha$, where the spacetime dimension is taken to be $D=5, 6, 7$, respectively, and the case of $D=4$ is attached for comparison. Here we set $M=1$.}
\label{figure4}
\end{figure}

\begin{figure}[h]
		\centering
		\begin{minipage}{.5\textwidth}
			\centering
			\includegraphics[width=80mm]{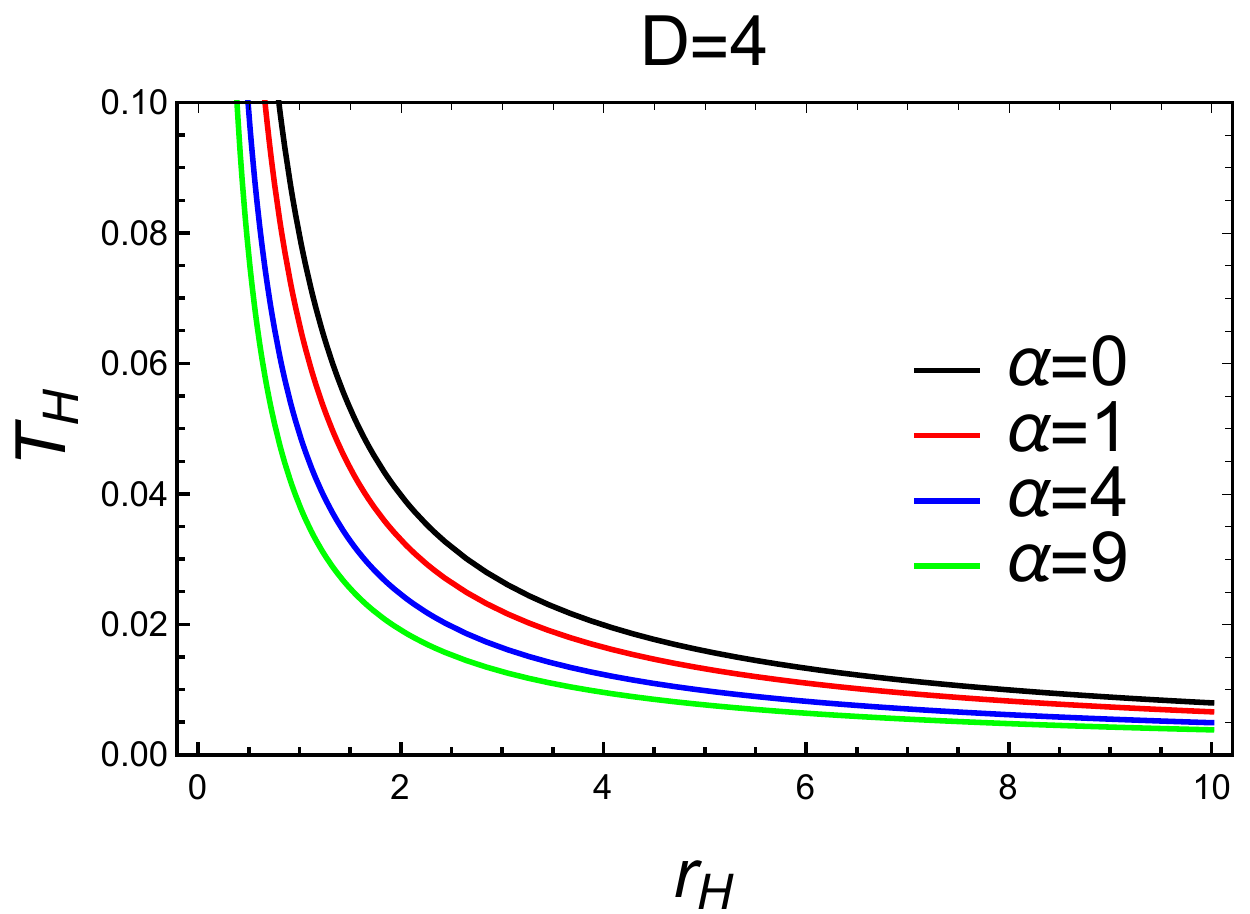}
		\end{minipage}%
		\begin{minipage}{.5\textwidth}
			\centering
			\includegraphics[width=80mm]{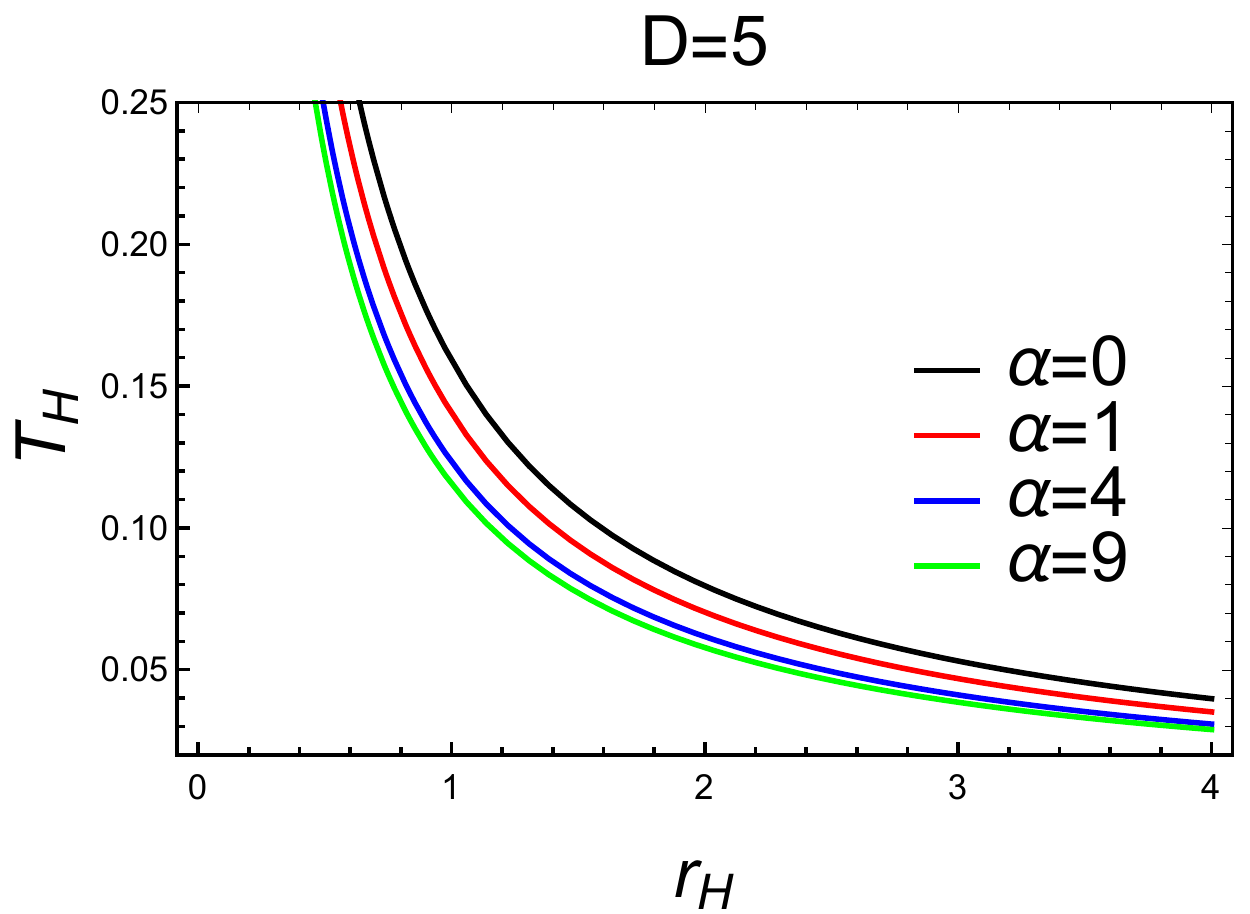}
		\end{minipage}
		\centering
		\begin{minipage}{.5\textwidth}
			\centering
			\includegraphics[width=80mm]{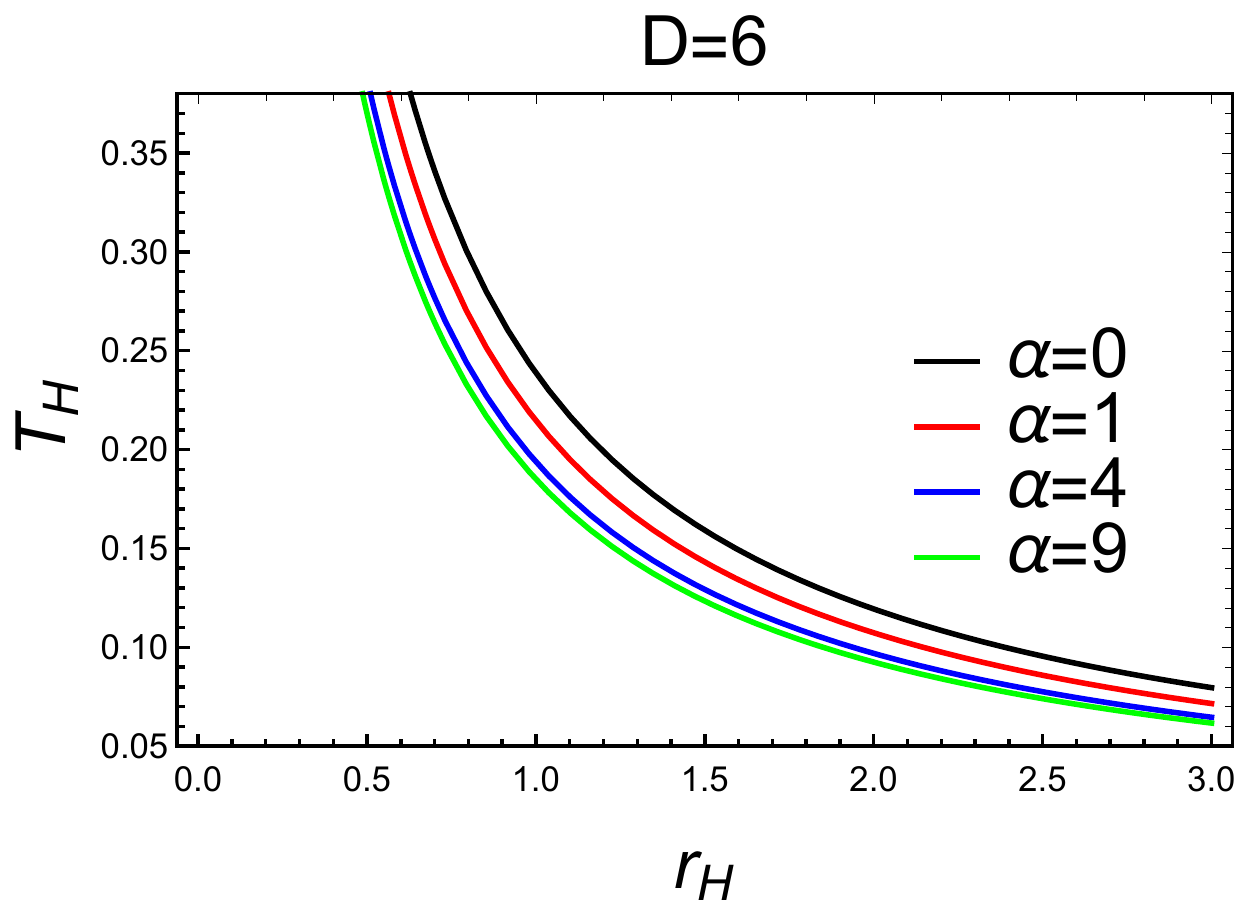}
		\end{minipage}%
		\begin{minipage}{.5\textwidth}
			\centering
			\includegraphics[width=80mm]{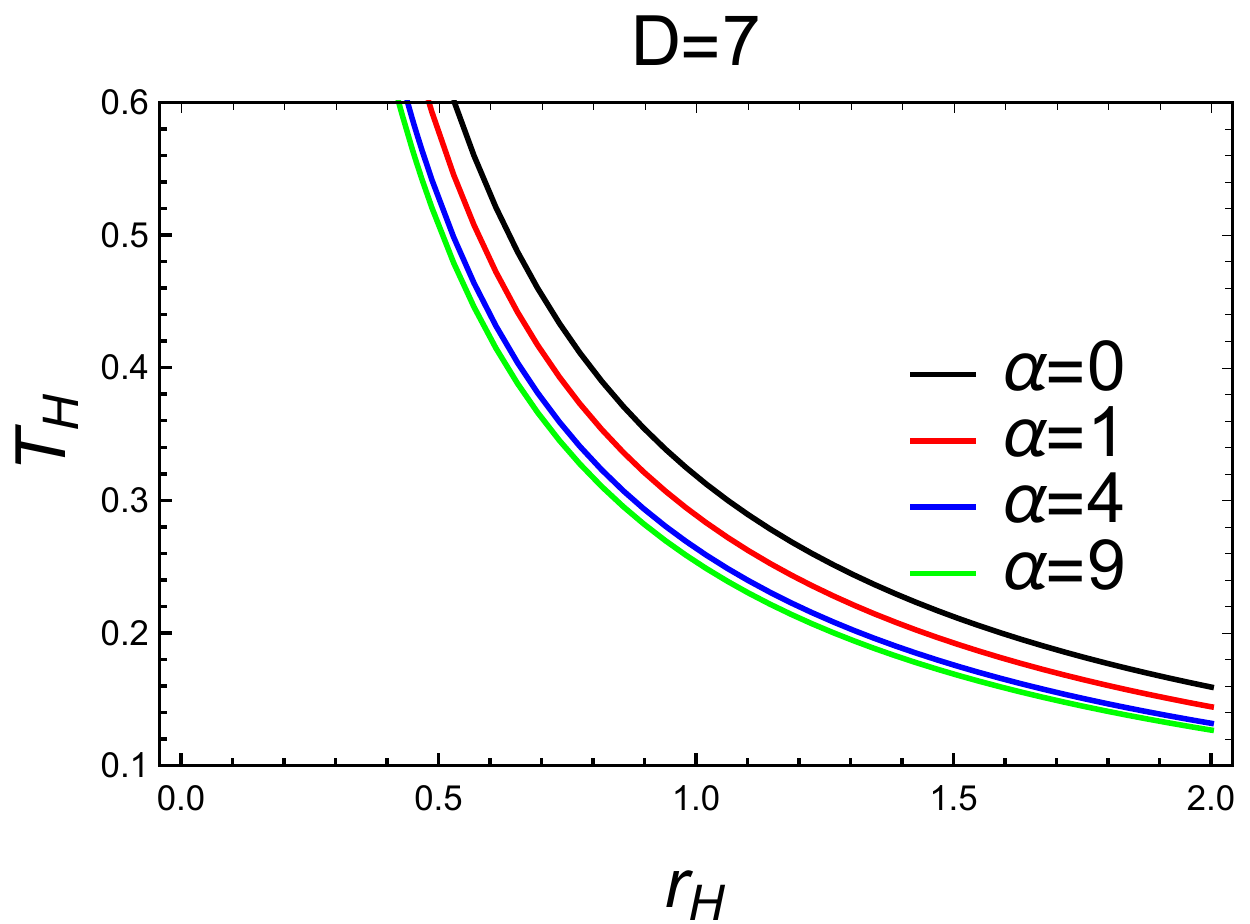}
		\end{minipage}
		\caption*{Fig. 2. Hawking temperature $T_{\rm H}$ with respect to $r_{\rm H}$ for different values of $\alpha$, where the spacetime dimension is taken to be $D=5, 6, 7$, respectively, and the case of $D=4$ is attached for comparison. }
\label{figure4}
\end{figure}

\section{Quasinormal mode frequencies of  massless scalar field perturbations for high-dimensional Schwarzschild  STVG black holes}

We consider a neutral massless scalar field perturbation around the high-dimensional Schwarzschild STVG black hole. The propagation of a neutral massless scalar field $\Phi$ in the curved black hole spacetime is described by the following Klein-Gordon equation~\cite{PC3},
\begin{equation}
\label{22}
\frac{1}{\sqrt{-g}}\partial _{\mu }(\sqrt{-g}g^{\mu \nu }\partial _{\nu }\Phi )=0,
\end{equation}
where $g$ is the determinant of the background metric tensor given by the line element Eq.~(\ref{10}).

By expanding the field into a spherical harmonic function, $\Phi =e^{-i\omega t}Y_{lm}(\chi){r^{-\frac{D-2}{2}}}{\Psi(r)}$, defining the tortoise coordinate, $dr_{*}=\frac{dr}{f(r)}$, and substituting the line element Eq.~(\ref{10}) into Eq.~(\ref{22}), we can finally get a radial perturbation equation as follows:
\begin{equation}
\label{23}
\frac{\mathrm{d} ^{2}\Psi (r)}{\mathrm{d} r^{2}_{*}}+[\omega ^{2}-V_{\rm eff}(r)]\Psi (r)=0,
\end{equation}
where $l$ is the multipole number, $\omega$ is the complex quasinormal mode frequency,  and $V_{\rm eff}(r)$  is the effective potential,
\begin{equation}
\label{24}
V_{\rm eff}(r)=f(r)\left[\frac{l(D+l-3)}{r^{2}}+\frac{(D-2)(D-4)f(r)}{4r^{2}}+\frac{(D-2)f'(r)}{2r}\right].
\end{equation}

\subsection{Quasinormal mode frequencies of  massless scalar field perturbations  calculated by the sixth-order WKB approximation}

Now we use the WKB approximation method to calculate numerically quasinormal mode frequencies of the massless  scalar field perturbation for the high-dimensional Schwarzschild STVG black hole. As to its usage, this method was first proposed by Schutz and Will~\cite{P11}, and then developed by Iyer and Will~\cite{P12} to the third order, by Konoplya~\cite{P13} to the sixth order, and by Matyjasek and Opala~\cite{P14} to the thirteenth order. We adopt the sixth-order WKB approximation for the sake of efficiency to calculate the quasinormal mode frequencies of the massless scalar field perturbation for the fundamental modes with the overtone number $n=0$ and multiple number $l=1, 2, 3$. The results are shown in Tables 1-4, where Table 1 is for the case of $D=4$. It should be noted  that the real and imaginary parts of quasinormal mode frequencies in the four tables represent the frequencies of the actual oscillation and the damping rates of the perturbation field, respectively.

\begin{table}[H]
\centering
\caption{The quasinormal mode frequencies of  massless scalar field perturbations calculated by the sixth-order WKB approximation for the four-dimensional Schwarzschild STVG black hole, where $\alpha=0$ corresponds to the case of Einstein's gravity.}
\label{table 1}
\begin{tabular}{|c|c|c|c|c|c|c|c|c|c|c|}
\hline
  \multicolumn{4}{|c|}{$D=4$, $M=1$}  \\ \hline

$\alpha $ &  $l=1$, $n=0$  &  $l=2$, $n=0$  &  $l=3$, $n=0$            \\  \hline
0&	0.29291 $-$ 0.0977616$i$	&		0.483642  $-$ 0.0967661$i$	&		0.675366  $-$ 0.0965006$i$		 \\ \hline
1&	0.161769  $-$ 0.0497066$i$	&		0.266908  $-$ 0.0492898$i$	&		0.372644  $-$ 0.0491757$i$		 \\ \hline
2&	0.112566 $-$ 0.0329813$i$	&		0.185707  $-$ 0.0327371$i$	&		0.25927  $-$ 0.03267$i$		       \\ \hline
3&	0.0864983 $-$ 0.024542$i$	&		0.14272 $-$ 0.0243782$i$&			0.199262 $-$ 0.0243329$i$		  \\ \hline
4&	0.0702981  $-$ 0.0194789$i$&		     0.116006  $-$ 0.0193616$i$&		     0.161974  $-$ 0.0193285$i$		  \\ \hline
5&	0.05923 $-$ 0.0161174$i$	&		0.0977626  $-$ 0.0160281$i$&		     0.13651  $-$ 0.0160024$i$		  \\ \hline
6&	0.0511861 $-$ 0.0137278$i$&		    0.0844999  $-$ 0.0136577$i$&		     0.117998 $-$ 0.0136369$i$		  \\ \hline
7&	0.0450687 $-$ 0.0119461$i$&		    0.0744174  $-$ 0.0118887$i$&		     0.103926  $-$ 0.0118713$i$		  \\ \hline
8&	0.0402549  $-$ 0.0105692$i$&		     0.0664913  $-$ 0.0105195$i$&		     0.0928617  $-$ 0.0105047$i$	        \\ \hline
9&	0.0363813  $-$ 0.00947074$i$&		0.0600948  $-$ 0.00942941$i$&		0.0839332  $-$ 0.00941643$i$	        \\ \hline
\end{tabular}
\end{table}

\begin{table}[H]
\centering
\caption{The quasinormal mode frequencies of massless  scalar field perturbations calculated by the sixth-order WKB approximation for the five-dimensional Schwarzschild STVG black hole, where $\alpha=0$ corresponds to the case of Einstein's gravity.}
\label{table 2}
\begin{tabular}{|c|c|c|c|c|c|c|c|c|c|c|}
\hline
  \multicolumn{4}{|c|}{$D=5$, $M=1$}  \\ \hline

$\alpha $ &  $l=1$, $n=0$  &  $l=2$, $n=0$  &  $l=3$, $n=0$            \\  \hline
0&	1.10108  $-$ 0.396434$i$	&		1.6395  $-$ 0.388248$i$	&		2.17936  $-$ 0.3862$i$		 \\ \hline
1&	0.799956 $-$ 0.274707$i$	&		1.18987 $-$ 0.269566$i$	&		1.58142  $-$ 0.268209$i$		 \\ \hline
2&	0.659853 $-$ 0.221695$i$	&		0.981074 $-$ 0.21788$i$	&		1.3039   $-$ 0.216847$i$		  \\ \hline
3&	0.574494 $-$ 0.190578$i$	&		0.854003 $-$ 0.187512$i$&	     1.13503  $-$ 0.186663$i$	 \\ \hline
4&	0.51551  $-$ 0.1696$i$	&		     0.766258 $-$ 0.16701$i$	&		1.01843  $-$ 0.166278$i$  \\ \hline
5&	0.471618 $-$ 0.15426$i$	&		     0.700992 $-$ 0.151997$i$	&	0.931702  $-$ 0.151347$i$		  \\ \hline
6&	0.437314 $-$ 0.142426$i$	&		0.649994 $-$ 0.140401$i$	&	0.863936  $-$ 0.139812$i$		  \\ \hline
7&	0.409547 $-$ 0.132944$i$	&		0.608723 $-$ 0.131099$i$	&	0.809093  $-$ 0.130558$i$		  \\ \hline
8&	0.386475 $-$ 0.125126$i$	&		0.574434 $-$ 0.123426$i$	&	0.763526  $-$ 0.122922$i$		  \\ \hline
9&	0.366909 $-$ 0.118539$i$	&		0.545355 $-$ 0.116955$i$	&	0.724884  $-$ 0.116483$i$		  \\ \hline

\end{tabular}
\end{table}

\begin{table}[H]
\centering
\caption{The quasinormal mode frequencies of massless  scalar field perturbations calculated by the sixth-order WKB approximation for the six-dimensional Schwarzschild STVG black hole, where $\alpha=0$ corresponds to the case of Einstein's gravity.}
\label{table 3}
\begin{tabular}{|c|c|c|c|c|c|c|c|c|c|c|}
\hline
  \multicolumn{4}{|c|}{$D=6$, $M=1$}  \\ \hline

$\alpha $ &  $l=1$, $n=0$  &  $l=2$, $n=0$  &  $l=3$, $n=0$            \\  \hline
0&	1.84019  $-$ 0.666898$i$&			2.57351  $-$ 0.642702$i$&			3.2999  $-$ 0.63815$i$		\\ \hline
1&	1.47989  $-$ 0.518071$i$&			2.06699  $-$ 0.500283$i$&			2.65059 $-$ 0.496464$i$		\\ \hline
2&	1.29927  $-$ 0.447346$i$&			1.81336  $-$ 0.433027$i$&			2.32541 $-$ 0.429844$i$		\\ \hline
3&	1.18343  $-$ 0.403642$i$&			1.6511   $-$ 0.391367$i$&			2.1174  $-$ 0.388581$i$		\\ \hline
4&	1.10023  $-$ 0.373028$i$&			1.53473  $-$ 0.362087$i$&			1.96823 $-$ 0.359568$i$		\\ \hline
5&	1.03635  $-$ 0.349934$i$&			1.44549  $-$ 0.339936$i$&			1.85383 $-$ 0.337612$i$		\\ \hline
6&	0.985111 $-$ 0.331643$i$&			1.37394  $-$ 0.322352$i$&			1.7621  $-$ 0.320176$i$		\\ \hline
7&	0.942694 $-$ 0.316647$i$&			1.31474  $-$ 0.307911$i$&			1.6862  $-$ 0.305853$i$		\\ \hline
8&	0.906746 $-$ 0.304034$i$&			1.26458  $-$ 0.295747$i$&			1.62189 $-$ 0.293785$i$		\\ \hline
9&	0.875718 $-$ 0.293212$i$&			1.22129  $-$ 0.285298$i$&			1.56639 $-$ 0.283419$i$		\\ \hline

\end{tabular}
\end{table}

\begin{table}[H]
\centering
\caption{The quasinormal mode frequencies of massless  scalar field perturbations calculated by the sixth-order WKB approximation for the seven-dimensional Schwarzschild STVG black hole, where $\alpha=0$ corresponds to the case of Einstein's gravity.}
\label{table 4}
\begin{tabular}{|c|c|c|c|c|c|c|c|c|c|c|}
\hline
  \multicolumn{4}{|c|}{$D=7$, $M=1$}  \\ \hline

$\alpha $ &  $l=1$, $n=0$  &  $l=2$, $n=0$  &  $l=3$, $n=0$            \\  \hline
0&	2.45993  $-$ 0.891478$i$&			3.3092   $-$ 0.837986$i$&			4.12768  $-$ 0.830792$i$	\\ \hline
1&	2.08458  $-$ 0.735942$i$&			2.80005  $-$ 0.693192$i$&			3.49467  $-$ 0.685619$i$	\\ \hline
2&	1.88976  $-$ 0.657704$i$&			2.53569  $-$ 0.621437$i$&			3.16515  $-$ 0.614671$i$	\\ \hline
3&	1.76137  $-$ 0.608152$i$&			2.36226  $-$ 0.575767$i$&			2.94893  $-$ 0.569573$i$	\\ \hline
4&	1.66727  $-$ 0.572814$i$&			2.23551  $-$ 0.543011$i$&			2.79087  $-$ 0.537225$i$	\\ \hline
5&	1.59386  $-$ 0.545762$i$&			2.13679  $-$ 0.517824$i$&			2.66775  $-$ 0.512345$i$	\\ \hline
6&	1.53418  $-$ 0.524065$i$&			2.05661  $-$ 0.497554$i$&			2.56774  $-$ 0.492318$i$	\\ \hline
7&	1.48421  $-$ 0.50608$i$&			1.98952  $-$ 0.480708$i$&			2.48405  $-$ 0.47567$i$      \\ \hline
8&	1.44144  $-$ 0.490805$i$&			1.93211  $-$ 0.466371$i$&			2.41243  $-$ 0.461499$i$	\\ \hline
9&	1.40419  $-$ 0.477584$i$&			1.88214  $-$ 0.453943$i$&			2.35008  $-$ 0.449212$i$	\\ \hline

\end{tabular}
\end{table}

In order to give an intuitive understanding for the data in Tables 1-4, we draw the graphs of real parts and negative imaginary parts of quasinormal mode frequencies with respect to parameter $\alpha$ in Fig. 3. It should be noted that the parameter $\alpha=0$ represents the case of the Schwarzschild-Tangherlini black hole of Einstein gravity. From this figure, we find that the increase of the parameter
$\alpha$ makes the real part and the negative imaginary part monotonically decrease in spacetime dimension $D=4, 5, 6, 7$, which  means that the larger the parameter $\alpha$  is, the more slowly the scalar  wave oscillates and decays in the high-dimensional Schwarzschild STVG black hole spacetime. However, the increase of spacetime dimension makes the real part and the negative imaginary part monotonically increase for a fixed $\alpha$ when the cases of $D=5, 6, 7$ are compared with the case of $D=4$, which  shows that  the larger the  spacetime dimension is, the faster the scalar  wave oscillates and  decays in the high-dimensional Schwarzschild STVG black hole spacetime.

\begin{figure}[H]
		\centering
		\begin{minipage}{.5\textwidth}
			\centering
			\includegraphics[width=80mm]{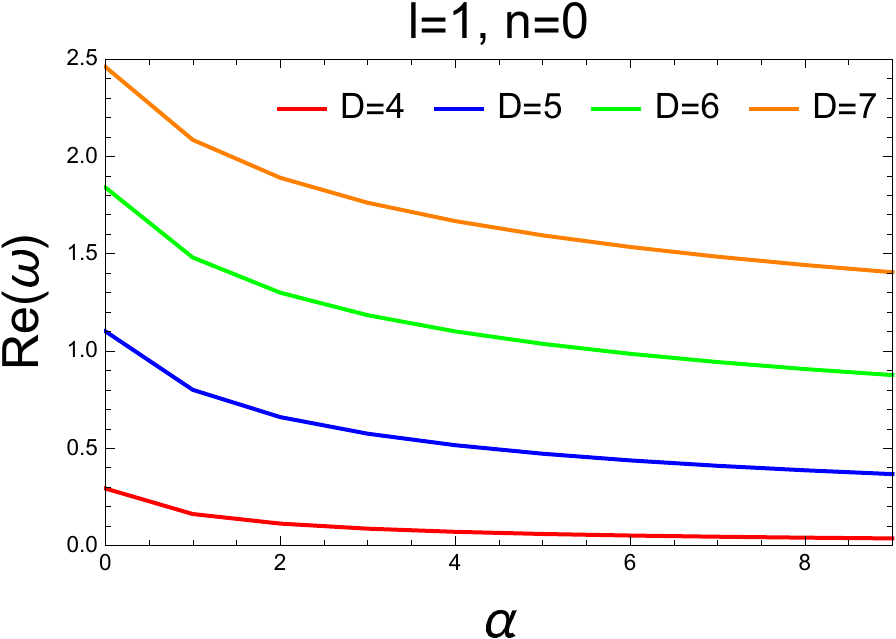}
		\end{minipage}%
		\begin{minipage}{.5\textwidth}
			\centering
			\includegraphics[width=80mm]{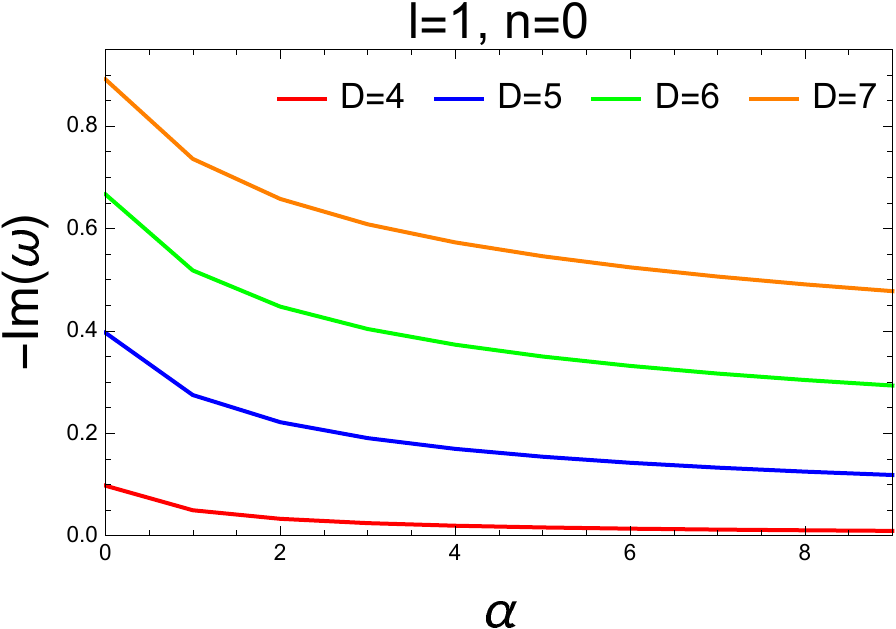}
		\end{minipage}
\end{figure}
\begin{figure}[H]
		\centering
		\begin{minipage}{.5\textwidth}
			\centering
			\includegraphics[width=80mm]{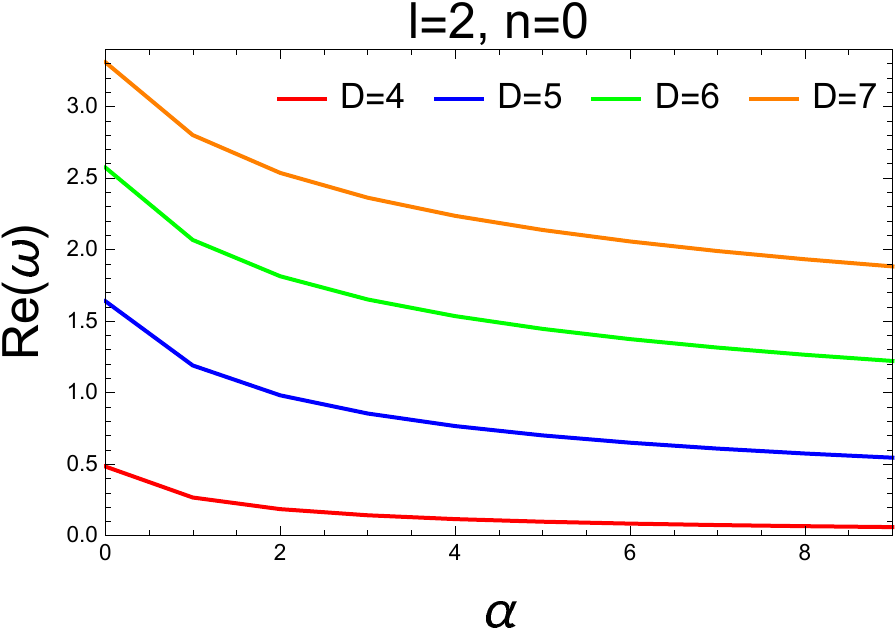}
		\end{minipage}%
		\begin{minipage}{.5\textwidth}
			\centering
			\includegraphics[width=80mm]{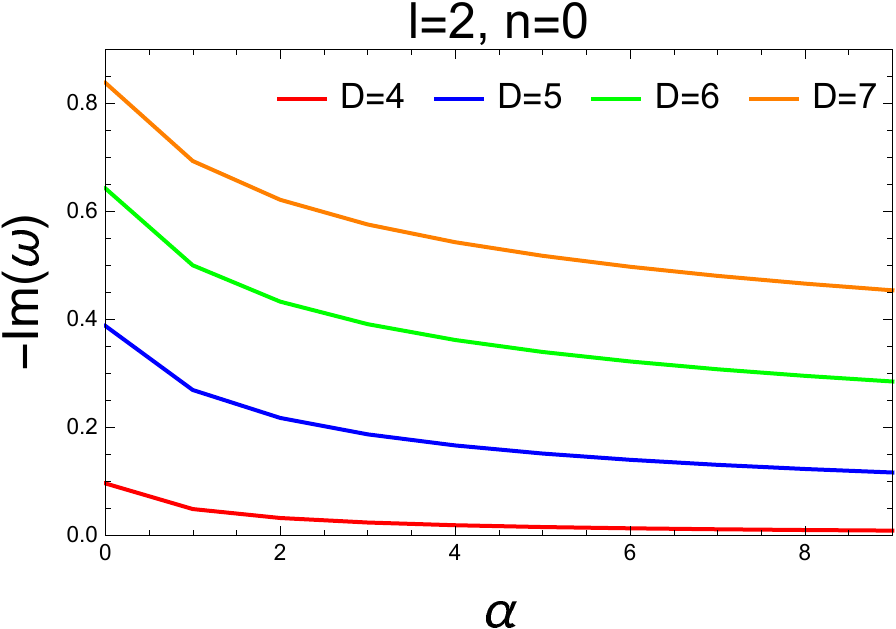}
		\end{minipage}
\end{figure}
\begin{figure}[H]
		\centering
		\begin{minipage}{.5\textwidth}
			\centering
			\includegraphics[width=80mm]{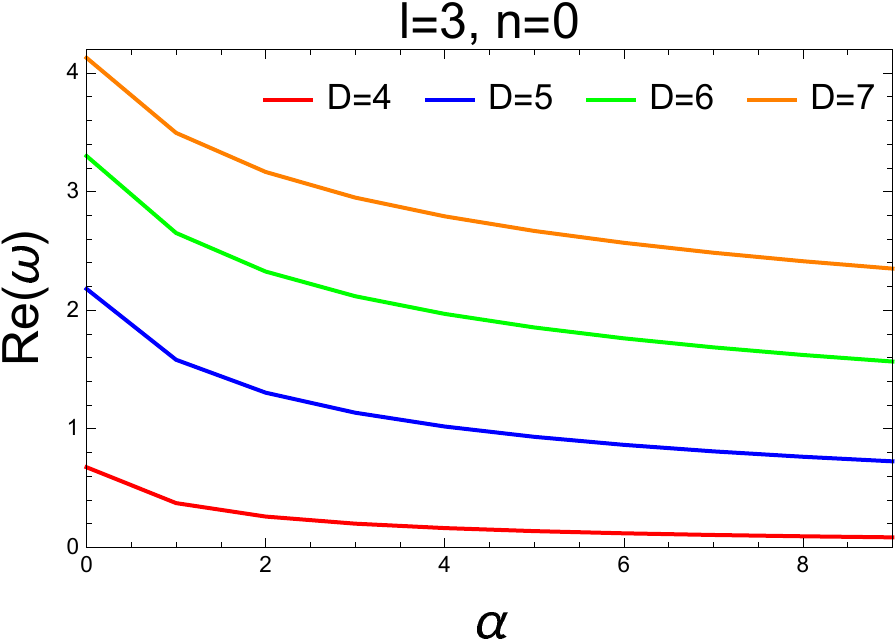}
		\end{minipage}%
		\begin{minipage}{.5\textwidth}
			\centering
			\includegraphics[width=80mm]{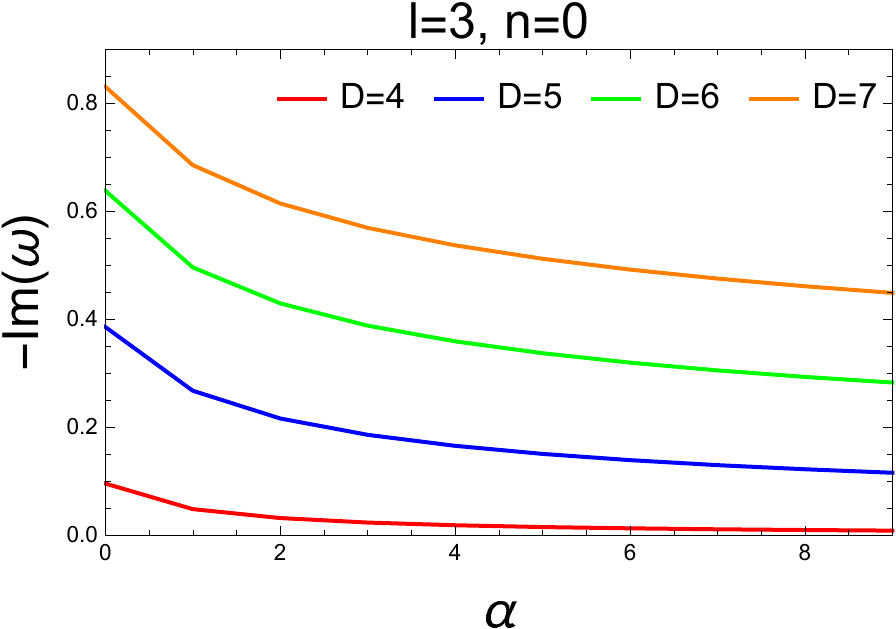}
		\end{minipage}
		\caption*{Fig. 3.  Real parts and negative imaginary parts of quasinormal frequencies of massless  scalar field perturbations with respect to $\alpha$, where the spacetime dimension is taken to be $D=5, 6, 7$, respectively, and the case of $D=4$ is attached for comparison. Here we set $M=1$.}
\label{figure4}
\end{figure}

\subsection{Quasinormal mode frequencies in the eikonal limit  calculated via unstable circular null geodesics}

The unstable circular null geodesic method used to calculate quasinormal mode frequencies of a static spherically symmetric black hole in the eikonal limit ($l\gg 1$) was first proposed~\cite{P15} by Cardoso {\em et al}.
The effective potential $V_{\rm eff}(r)$, see Eq.~(\ref{24}), takes the following form under this limit,
\begin{equation}
\label{25}
V(r)=f(r)\frac{l^{2}}{r^{2}}+O(l).
\end{equation}
The quasinormal mode frequencies $\omega $ can be calculated by the formula,
\begin{equation}
\label{26}
\omega _{l\gg 1}=l\Omega_c -i\left(n+\frac{1}{2}\right)|\lambda_{\rm L}|,
\end{equation}
with
\begin{equation}
\label{27}
\Omega _c=\frac{\sqrt{f(r_{c})}}{r_{c}},        \qquad   \lambda_{\rm L} =\sqrt{\frac{f(r_{c})(2f(r_{c})-r_{c}^{2}{f}''(r_{c}))}{2r_{c}^{2}}},
\end{equation}
where $\Omega_c$ denotes the angular velocity and $\lambda_{\rm L}$ the Lyapunov exponent which determine the real and imaginary parts of quasinormal mode frequencies, respectively.
In addition, $r_{c}$ stands for the radius of unstable circular null geodesics determined by
\begin{equation}
\label{28}
2f(r_{c})-r_{c}\left.\frac{\mathrm{d} f(r)}{\mathrm{d} r}\right|_{r=r_{c}}=0.
\end{equation}

Recently, the relation between real parts of quasinormal frequencies in the eikonal limit and shadow radii of black holes has been gained~\cite{P16}, i.e., ${\rm Re}\,\omega _{l\gg 1}=\frac{l}{R_{\rm sh}}$, where
\begin{equation}
R_{\rm sh}=\frac{r_{c}}{\sqrt{f(r_{c})}}\label{shadow}
\end{equation}
represents the shadow radius of black holes and equals the inverse of the angular velocity $\Omega_c $ in the eikonal limit. Subsequently, this relation was improved~\cite{P17} for a high-dimensional static spherically symmetric black hole in terms of the WKB method, i.e., ${\rm Re}\,\omega _{l\gg 1}=\frac{l+\frac{D-3}{2}}{R_{\rm sh}}$.  Therefore, the improved expression of the quasinormal mode frequencies in the eikonal limit for a high-dimensional static spherically symmetric black hole takes the form,
\begin{equation}
\label{29}
 \omega _{l\gg 1}=\frac{l+\frac{D-3}{2}}{R_{\rm sh}}-i\left(n+\frac{1}{2}\right)|\lambda_{\rm L}|=\left(l+\frac{D-3}{2}\right)\Omega_c -i\left(n+\frac{1}{2}\right)|\lambda_{\rm L}|.
\end{equation}

By substituting the metric function of the high-dimensional Schwarzschild STVG black hole, Eq.~(\ref{13}), into
Eqs.~(\ref{27}) and ~(\ref{28}), we can obtain the angular velocity $\Omega_c $ and the Lyapunov exponent $\lambda_{\rm L}$. Then we can plot the graph of the angular velocity $\Omega_c$  and the Lyapunov exponent $\lambda_{\rm L}$ with respect to parameter $\alpha$ as shown in Fig. 4. We find from this figure that the increase of parameter $\alpha$ makes the angular velocity $\Omega_c$ and the Lyapunov exponent $\lambda_{\rm L}$ monotonically decrease in spacetime dimension $D=4, 5, 6, 7$, which means the increase of parameter $\alpha$ makes real parts of quasinormal frequencies in the eikonal limit  and  negative imaginary parts monotonically decrease. However, the increase of spacetime dimension makes the angular velocity $\Omega_c$ and the Lyapunov exponent $\lambda_{\rm L}$ monotonically increase for a fixed $\alpha$ when the cases of $D=5, 6, 7$ are compared with the case of $D=4$,  which shows  that  the increase of spacetime dimension makes  real parts of quasinormal frequencies in the eikonal limit  and  negative imaginary parts  monotonically increase.

\begin{figure}[H]
		\centering
		\begin{minipage}{.5\textwidth}
			\centering
			\includegraphics[width=80mm]{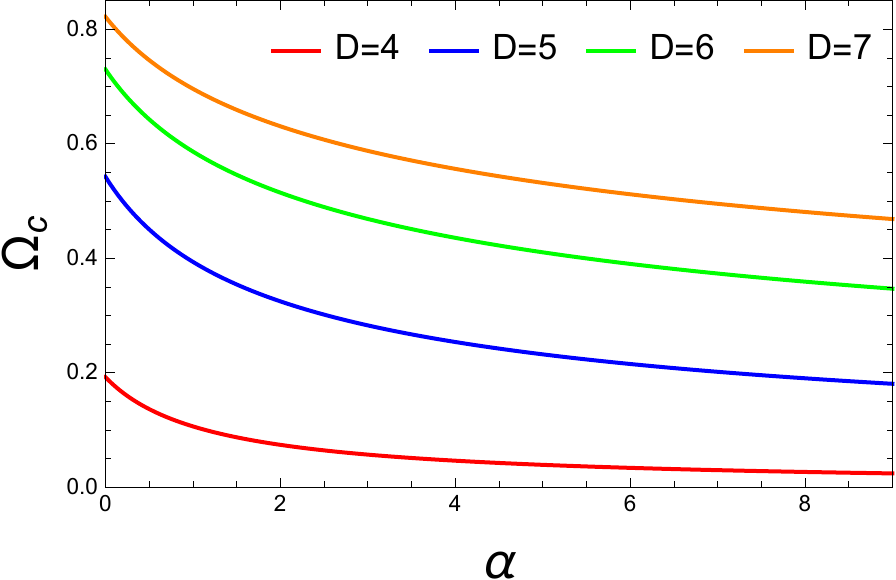}
		\end{minipage}%
		\begin{minipage}{.5\textwidth}
			\centering
			\includegraphics[width=80mm]{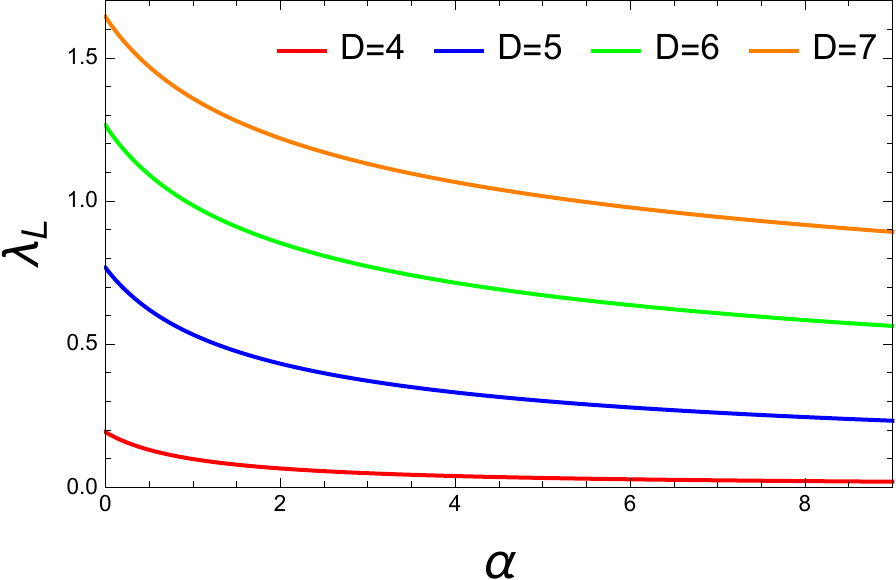}
		\end{minipage}
		\caption*{Fig. 4.  The angular velocity $\Omega_c$ and the Lyapunov exponent $\lambda_{\rm L}$  with respect to the parameter $\alpha$, where the spacetime dimension is taken to be $D=5, 6, 7$, respectively, and the case of $D=4$ is attached for comparison. Here we set $M=1$.}
\label{figure4}
\end{figure}

Therefore, based  on  the comparison and analysis  of  Fig. 3 and  Fig. 4, we can give the following conclusions:

\begin{itemize}

\item  The larger the parameter $\alpha$  is, the more slowly the scalar  wave oscillates and decays in the high-dimensional Schwarzschild STVG black hole spacetime.

\item  The larger the  spacetime dimension is, the faster the scalar  wave oscillates and  decays in the high-dimensional Schwarzschild STVG black hole spacetime.

\item  The spacetime of the high-dimensional Schwarzschild STVG black hole  is more stable than that of the Schwarzschild-Tangherlini black hole.

\end{itemize}

\section{Shadow  radius  and  energy emission rate of high-dimensional \\Schwarzschild STVG black holes}

In this section, we first study the shadow radius $R_{\rm sh}$ and then use it to calculate the energy emission rate for the high-dimensional Schwarzschild STVG black hole.

By using the data of the shadow radius obtained in subsec. 4.2, see Eq.~(\ref{shadow}), we draw the graph of the shadow radius $R_{\rm sh}$ with respect to parameter $\alpha$ in Fig. 5, where the spacetime dimension is taken to be $D=5, 6, 7$, respectively, and the case of $D=4$ is attached. We find that the increase of parameter $\alpha$ makes the shadow radius increase, while the increase of the spacetime dimension makes the shadow radius decrease, which is consistent with the relevant conclusions in the four-dimensional Schwarzschild STVG black hole~\cite{PP3, P36} and the high-dimensional Schwarzschild-Tangherlini black hole~\cite{P37}.  We notice the significant difference between the shadow radius $R_{\rm sh}$ of the four-dimensional Schwarzschild STVG black hole and that of the high-dimensional ones ($D=5, 6, 7$) in Fig. 5 for a fixed $\alpha$.

\begin{figure}[h]
		\centering
		\begin{minipage}{.5\textwidth}
			\centering
			\includegraphics[width=80mm]{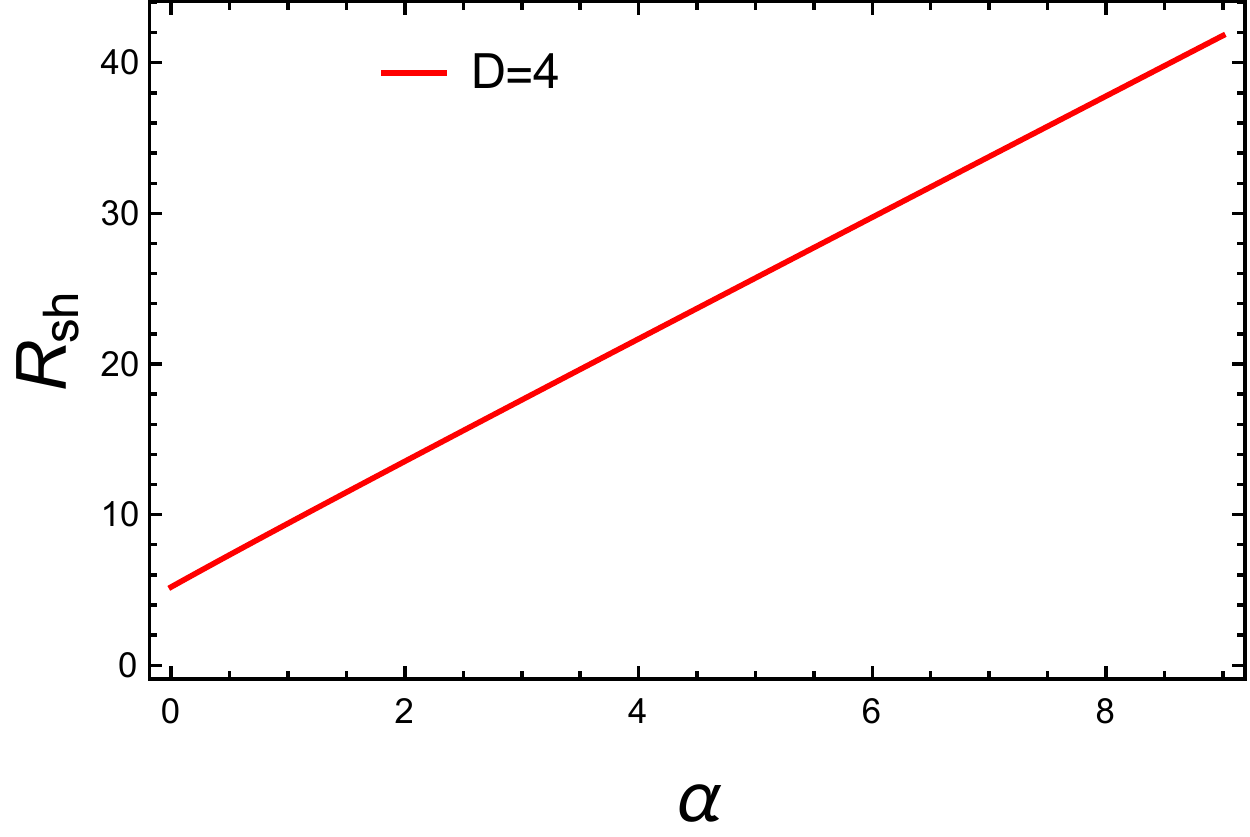}
		\end{minipage}%
		\begin{minipage}{.5\textwidth}
			\centering
			\includegraphics[width=80mm]{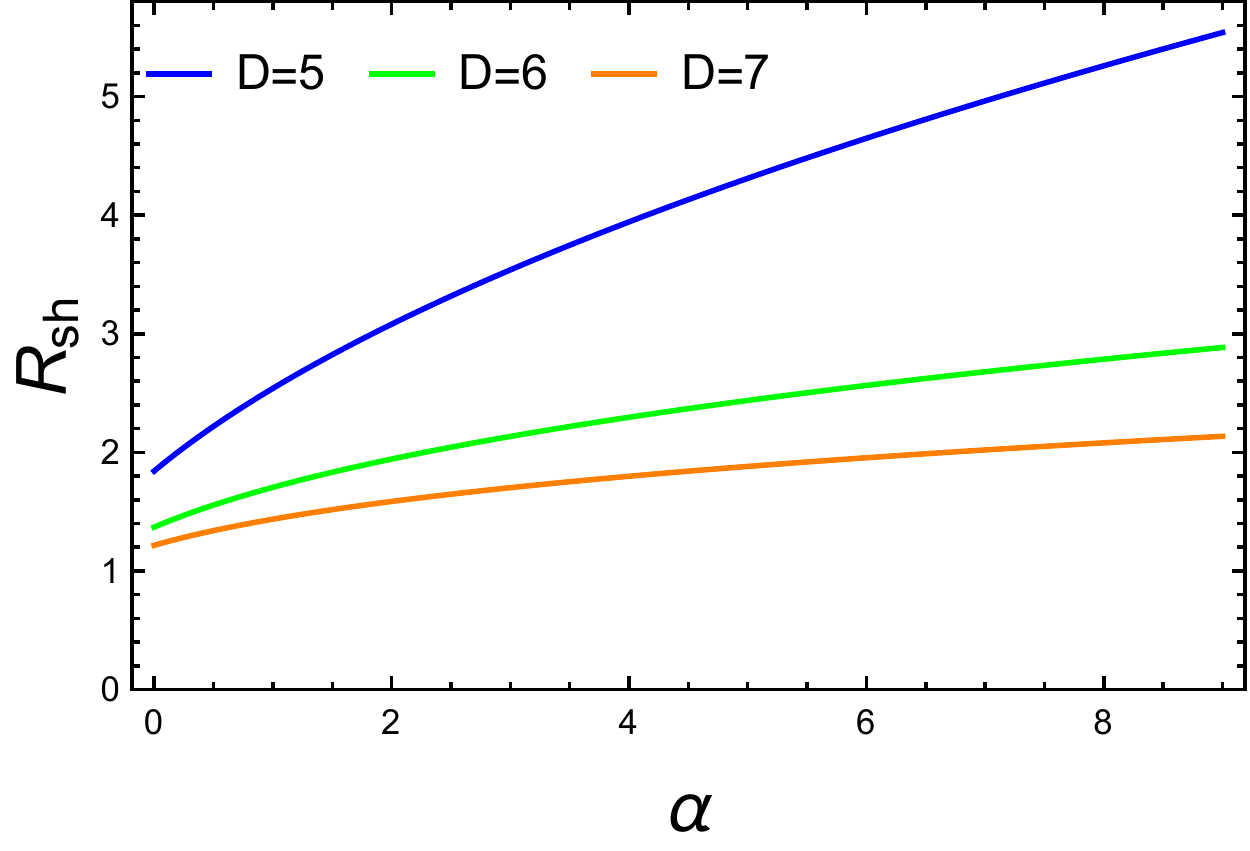}
		\end{minipage}
		\caption*{Fig. 5.  The shadow radius $R_{\rm sh}$ with respect to parameter $\alpha$, where the spacetime dimension is taken to be $D=5, 6, 7$, respectively, and the case of $D=4$ (left diagram) is attached for comparison. Here we set $M=1$.}
\label{figure4}
\end{figure}

The high-energy absorption cross section in a spherically symmetric black hole spacetime oscillates~\cite{PP5} around a limiting constant $\sigma_{\rm lim}$ which is approximately related~\cite{PP6} to the black hole shadow as follows,
\begin{equation}
\label{31}
\sigma _{\rm lim}\approx \frac{\pi ^{\frac{D-2}{2}}R^{D-2}_{\rm sh}}{\Gamma (\frac{D}{2})}.
\end{equation}
Therefore, we can use the shadow radius to calculate the energy emission rate of the high-dimensional Schwarzschild STVG black hole using the formula~\cite{PP6,PP8, PP7},
\begin{equation}
\label{30}
\frac{\mathrm{d}^{2} E(\varpi )}{\mathrm{d}\varpi \mathrm{d}t}=2\pi ^{2}\sigma _{\rm lim}\,\frac{\varpi ^{D-1}}{\exp({\varpi}/{T_{\rm H}})-1},
\end{equation}
where $\varpi$ represents the emission frequency, $T_{\rm H}$ the Hawking temperature, and $R_{\rm sh}$ the shadow radius, see Eq.~(\ref{21}) and Eq.~(\ref{shadow}).

We plot the graph of the energy emission rate $\frac{\mathrm{d}^{2} E(\varpi )}{\mathrm{d}\varpi \mathrm{d}t}$ with respect to the emission frequency $\varpi$ for different values of $\alpha$ in spacetime dimension $D=5, 6, 7$ in Fig. 6, where the case of $D=4$, see Refs.~\cite{PP3,P2}, is attached for comparison. It should be noted that the parameter $\alpha=0$ represents the case of the Schwarzschild-Tangherlini black hole of Einstein's gravity. On the one hand, we can see that when $D$ is a given high-dimensional spacetime dimension from five to seven, the peak of the energy emission rate decreases as the parameter $\alpha$ increases in the high-dimensional Schwarzschild STVG black hole, which is similar to the situation in the four-dimensional Schwarzschild STVG black hole~\cite{PP3}. On the other hand, we can clearly see that the peak of the energy emission rate increases rapidly with the increase of the spacetime dimension for a fixed $\alpha$, which is similar to situation in the Schwarzschild-Tangherlini black hole~\cite{PP4}. Therefore, we deduce such a conclusion: for the high-dimensional Schwarzschild STVG black hole, the increase of parameter $\alpha$ makes the black hole evaporation process slow, while the increase of the spacetime dimension makes this process fast.

\begin{figure}[H]
		\centering
		\begin{minipage}{.5\textwidth}
			\centering
			\includegraphics[width=80mm]{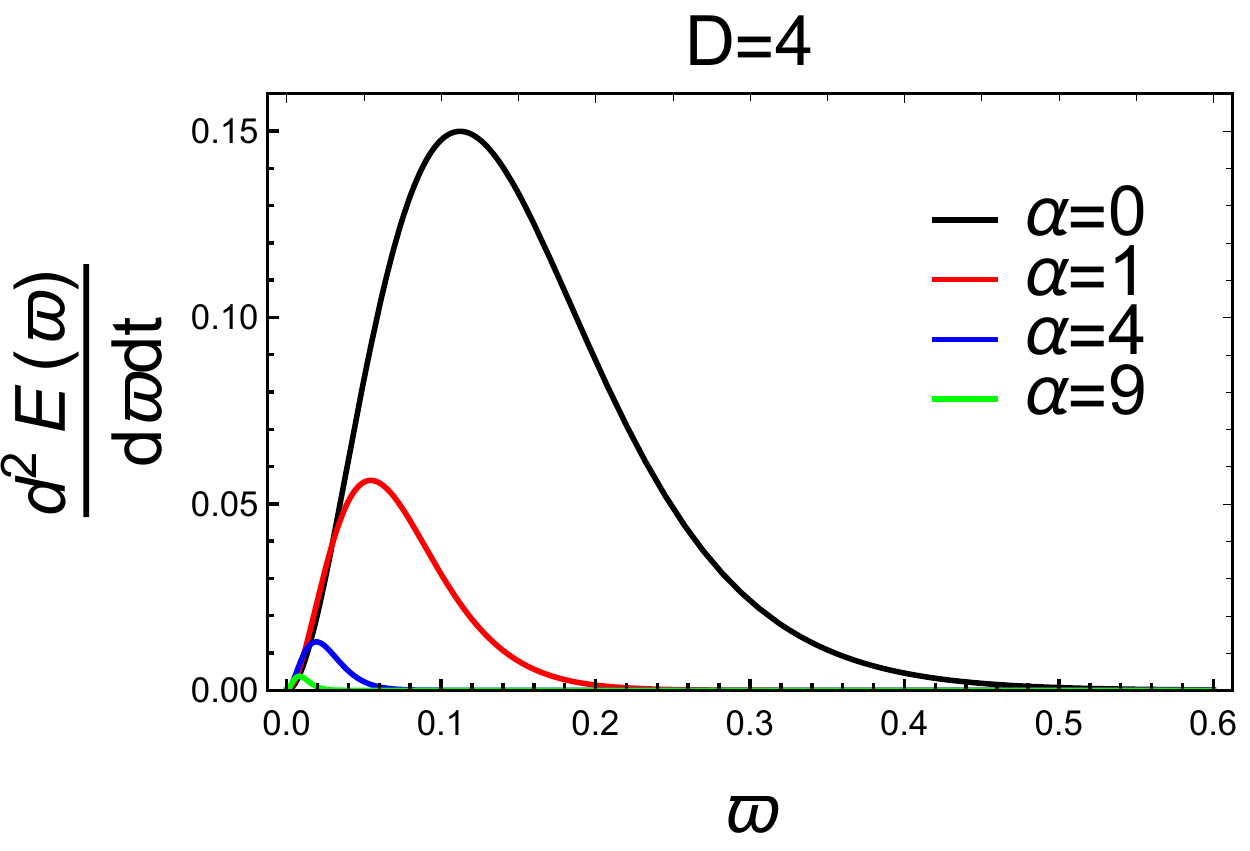}
		\end{minipage}%
		\begin{minipage}{.5\textwidth}
			\centering
			\includegraphics[width=80mm]{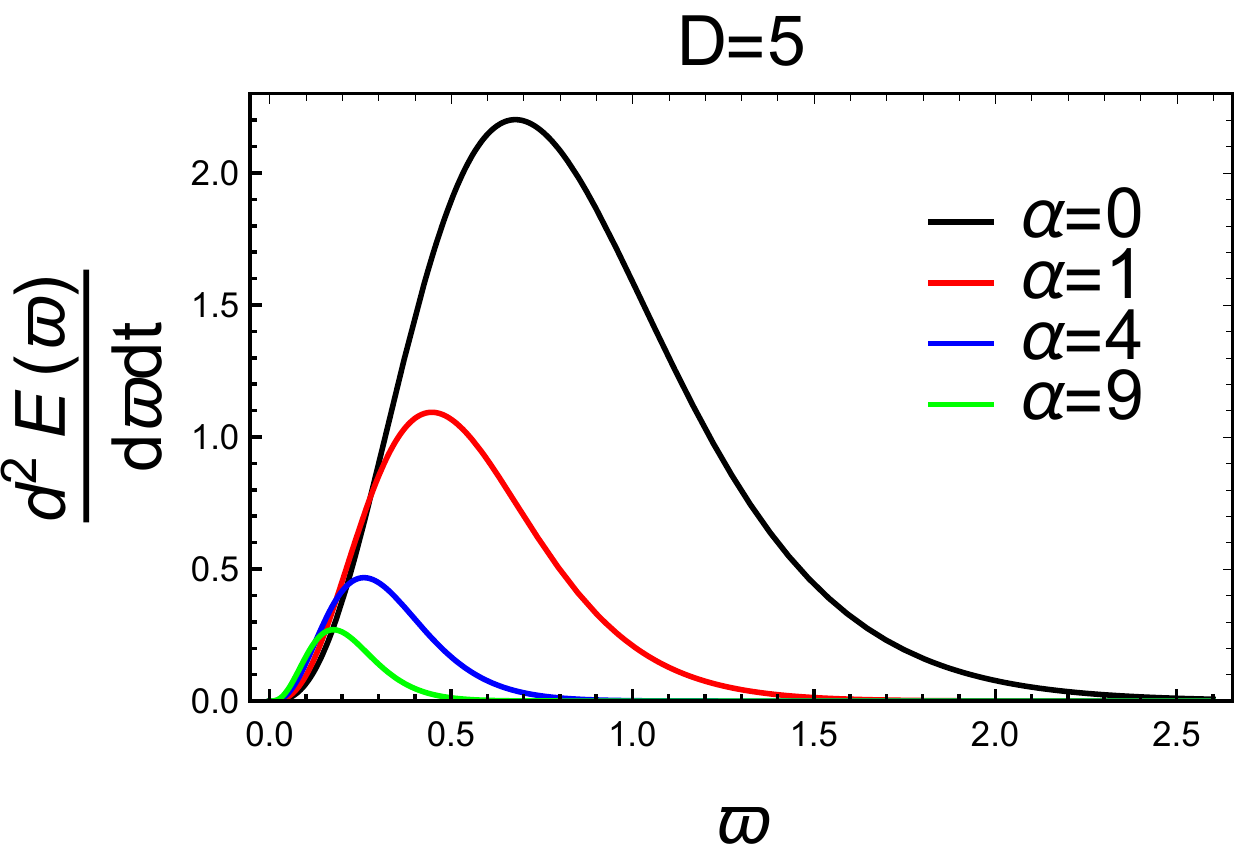}
		\end{minipage}
\end{figure}
\begin{figure}[H]
		\centering
		\begin{minipage}{.5\textwidth}
			\centering
			\includegraphics[width=80mm]{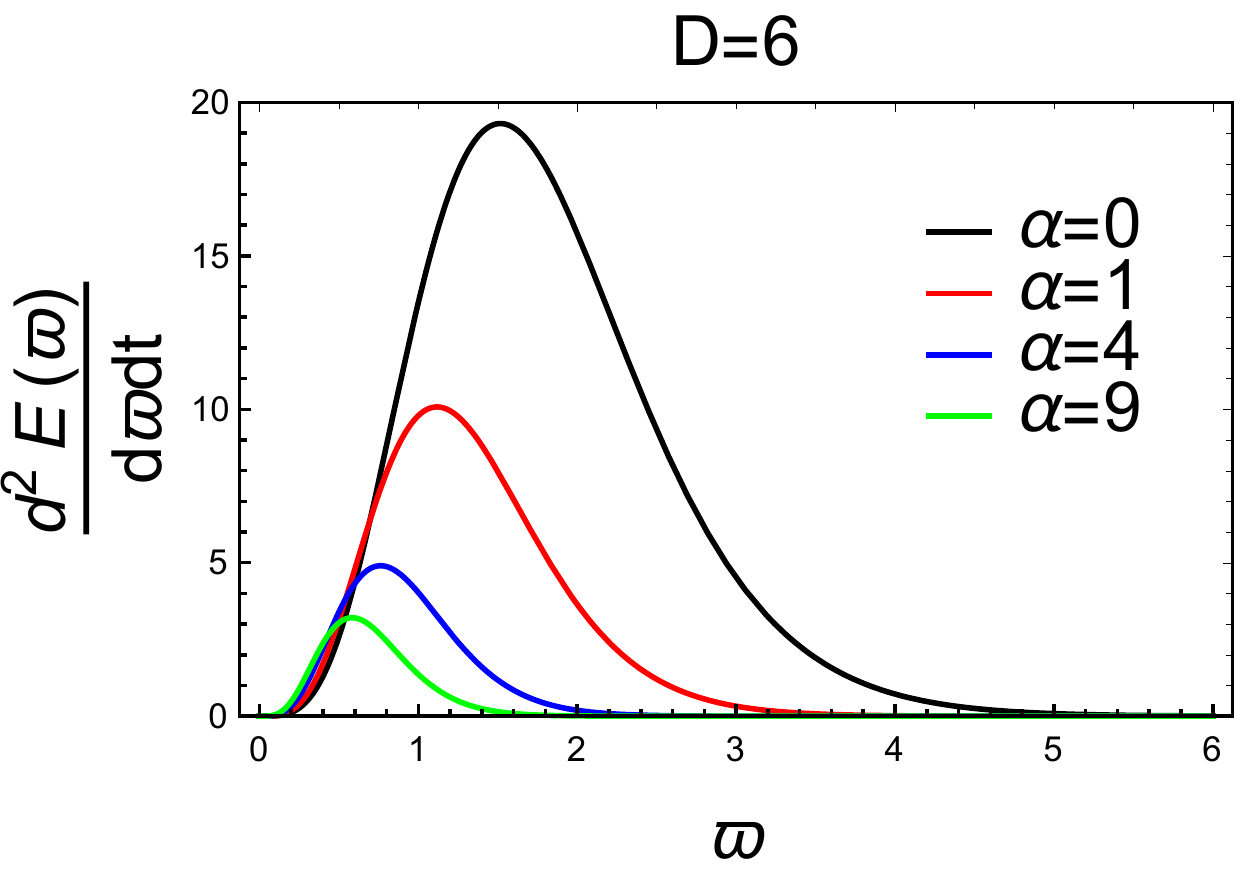}
		\end{minipage}%
		\begin{minipage}{.5\textwidth}
			\centering
			\includegraphics[width=80mm]{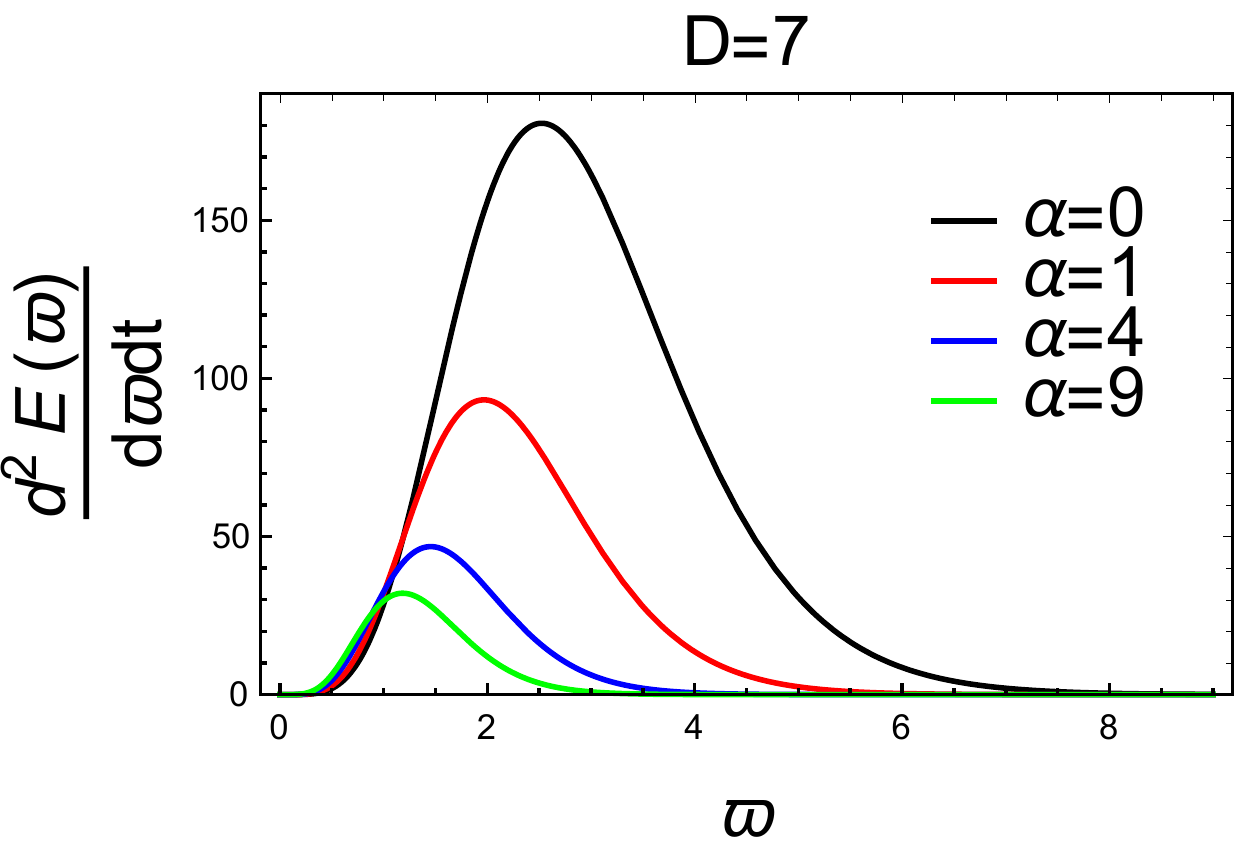}
		\end{minipage}
		\caption*{Fig. 6.  The energy emission rate $\frac{\mathrm{d}^{2} E(\varpi )}{\mathrm{d}\varpi \mathrm{d}t}$  with respect to the emission frequency $\varpi$ for different values of  $\alpha$, where the spacetime dimension is taken to be $D=5, 6, 7$, respectively, and the case of $D=4$ is attached for comparison. Here we set $M=1$.}
\label{figure4}
\end{figure}

\section{Conclusion }

In this paper, we first obtain a high-dimensional Schwarzschild STVG black hole solution, and then analyze the influence of parameter $\alpha$ on the event horizon radius and Hawking temperature. We find that the increase of parameter $\alpha$ makes the event horizon radius increase and the Hawking temperature decrease. We calculate quasinormal mode frequencies of massless scalar field perturbations for the high-dimensional Schwarzschild black hole using the sixth-order WKB approximation. We plot the graphs of real parts and negative imaginary parts of quasinormal frequencies of massless scalar field perturbations with respect to parameter $\alpha$.  We also use the unstable null geodesic method to compute the quasinormal mode frequencies in the eikonal limit and plot the angular velocity $\Omega_c$ and the Lyapunov exponent $\lambda_{\rm L}$ with respect to parameter $\alpha$.
The results show that the increase of parameter $\alpha$ makes the scalar waves decay slowly, while the increase of the spacetime dimension makes the scalar waves decay fast in the  high-dimensional Schwarzschild STVG black hole spacetime. In addition, we draw a graph of the shadow radius $R_{\rm sh}$ with respect to parameter $\alpha$ and find that the increase of parameter $\alpha$ makes the shadow radius of the high-dimensional Schwarzschild STVG black hole increase, while the increase of the spacetime dimension makes the black hole shadow radius decrease.  Finally, we investigate the energy emission rate of the high-dimensional Schwarzschild STVG black hole, and find that the increase of parameter $\alpha$ makes the evaporation process slow, while the increase of the spacetime dimension makes this process fast.

\section*{Acknowledgments}

The authors would like to thank  C. Lan for helpful discussions.
This work was supported in part by the National Natural Science Foundation of China under Grant No. 11675081. The authors would like to thank the anonymous referee for the helpful comments that improve this work greatly.

\end{document}